\documentclass[twocolumn,showpacs,amsfonts]{revtex4-1}
\usepackage{amsmath}
\usepackage{latexsym}
\usepackage{float}
\usepackage{amssymb}
\usepackage{graphicx}
\usepackage{color}
\usepackage{textcomp}
\usepackage{hyperref}
\usepackage{geometry}

 1

\def\ba{\begin{eqnarray}}
\def\ea{\end{eqnarray}}
\def\be{\begin{equation}}
\def\ee{\end{equation}}
\def\bm{\begin{math}}
\def\me{\end{math}}

\newcommand{\dummy}

\begin{document}
\title{Finite-size scaling study of dynamic critical phenomena in a vapor-liquid transition}
\author{Jiarul Midya and Subir K. Das$^{*}$}
\affiliation{Theoretical Sciences Unit, Jawaharlal Nehru Centre for Advanced Scientific Research,
 Jakkur P.O., Bangalore 560064, India}

\date{\today}

\begin{abstract}
~Via a combination of molecular dynamics (MD) simulations and finite-size scaling (FSS) analysis, we study dynamic
critical phenomena for the vapor-liquid transition in a three dimensional Lennard-Jones system. The phase behavior 
of the model has been obtained via the Monte Carlo simulations. The transport 
properties, viz., the bulk viscosity and the thermal conductivity, are calculated via the Green-Kubo relations, by taking 
inputs from the MD simulations in the microcanonical ensemble. The critical singularities of these quantities are estimated via 
the FSS method. The results thus obtained are in nice agreement with the predictions of the dynamic renormalization group 
and mode-coupling theories.
\end{abstract} 


\maketitle
\section{Introduction}
~ Understanding of the anomalous behavior of various static and dynamic quantities, in the vicinity 
of the critical points
\cite{2MEFisher1,2MEFisher2,2MEFisher3,2HEStanley,2PCHohenberg,2AOnuki1,2VPrivman,rev_sengers,2DPLandau,
2MAnisimov,2LMistura,2HCBurstyn1,2RAFerrell1,
2HCBurstyn2,2RAFerrell2,2GAOlchowy,2RFolk1,2JLStrathmann,2AOnuki2,2RFolk2,2JZJustin,2HHao,2JKBhattacharjee1,2JKBhattacharjee2},
is of fundamental importance. The critical behavior of the static quantities have been 
understood to a good extent via analytical theories, 
experiments and computer simulations \cite{2MEFisher1,2MEFisher2,2MEFisher3,2HEStanley,2AOnuki1,2VPrivman,2DPLandau}. 
On the other hand, the situation with respect to dynamics is relatively poor. Simulation studies, that helped achieving 
the objective for the static phenomena, gained momentum in the context of dynamic critical phenomena only 
recently \cite{2KJagannathan1,2KMeier,2KJagannathan2,2AChen,2SKDas1,2SKDas2,2KDyer,2SRoy1,2MGross,2SRoy2,2SRoy3,2SRoy4,2SRoy5,2JWMutoru}. 
Such a status is despite the fact that adequate information on the equilibrium transport phenomena  
is very much essential for the understanding of even nonequilibrium phenomena like the kinetics of 
phase transitions \cite{2AOnuki1,2AJBray}. For example, the crossovers and amplitudes in the growth-laws 
during phase transitions are often directly connected to the quantities like diffusivity and viscosity \cite{2AJBray,2HFurukawa}.  
\par
~ The static correlation length, $\xi$, diverges at the critical point \cite{2HEStanley}, i.e., 
$\xi \rightarrow \infty$ as the temperature $T \rightarrow T_c$, $T_c$ being the critical point value for the latter. 
As a result, various other static as well as dynamics quantities show singularities in approach to the criticality. 
These singularities are of power-law type, in terms of the reduced temperature ($\epsilon =|T-T_c|/T_c$), 
such as \cite{2MEFisher1,2MEFisher2,2MEFisher3,2HEStanley,2AOnuki1}
\begin{equation}\label{chap2_static_quan}
 \xi \sim \epsilon^{-\nu}, ~~\psi \sim \epsilon^{\beta}, ~~C \sim \epsilon^{-\alpha}, ~~\chi \sim \epsilon^{-\gamma}.
\end{equation}
Here, $\psi$, $C$ and $\chi$ are the order-parameter, specific heat and susceptibility, respectively. Typically, 
singularities for various dynamic quantities, viz., mutual or thermal diffusivity ($D$), shear viscosity ($\eta$), 
bulk viscosity ($\zeta$), thermal conductivity ($\lambda$), etc., are expressed in 
terms of $\xi$ as \cite{2PCHohenberg,2VPrivman,2MAnisimov} 
\begin{equation}\label{chap2_dynamic_quan}
 D \sim \xi^{-x_D}, ~~\eta \sim \xi^{x_{_{\eta}}}, ~~\zeta \sim \xi^{x_{_{\zeta}}}, ~~\lambda \sim \xi^{x_{_{\lambda}}}.
\end{equation}
\par
~ The static critical exponents do not depend upon the choice of material and the type of transition. 
In a particular dimension ($d$), if the interaction among the particles or spins are of same type, 
i.e., either of short or long range, and the order parameters have the same number of components, the exponents 
will have the same values, giving rise to well defined universality classes. 
For short range interactions with one component order-parameters, the exponents belong to the Ising universality 
class \cite{2MEFisher1,2MEFisher2,2MEFisher3,2HEStanley,2JZJustin}. The universality of the critical exponents 
in statics, thus, is very robust, viz., paramagnetic to ferromagnetic, 
liquid-liquid, vapor-liquid transitions will all have the same set of exponent values depending upon the interaction range.   
Values of the above mentioned static exponents for the $d=3$ Ising class are \cite{2JZJustin}
\begin{equation}\label{chap2_static_exp_val}
\nu \simeq 0.63, ~~\alpha \simeq 0.11,~~\beta \simeq 0.325,~~\gamma \simeq 1.239.
\end{equation}
\par
~ On the other hand, the universality of the dynamic exponents is considerably weaker. For example, the value of  
the exponent $z$, related to the longest relaxation time \cite{2DPLandau}
\begin{equation}
\tau \sim \xi^{z}, 
\end{equation}
can vary depending upon the choice of statistical 
ensemble \cite{2PCHohenberg,2AOnuki1,2DPLandau}. Nevertheless, the exponents for liquid-liquid and vapor-liquid transitions 
should be same, given by the fluid or model H universality class \cite{2PCHohenberg,2AOnuki1,rev_sengers}. 
The values of these exponents for this class are
\begin{equation}\label{chap2_dyna_exp_val}
x_{_{\lambda}}\simeq 0.902, ~~x_{_{\eta}} \simeq 0.068,~~x_{_{\zeta}} \simeq 2.893,~~x_{_{D}} \simeq 1.068.
\end{equation} 
These numbers are obtained via the dynamic renormalization group and mode-coupling theoretical calculations and 
found to be in agreement with experiments 
\cite{2PCHohenberg,2AOnuki1,2MAnisimov,2LMistura,2HCBurstyn1,2RAFerrell1,2HCBurstyn2,2RAFerrell2,2GAOlchowy,
2RFolk1,2JLStrathmann,2AOnuki2,2RFolk2,2HHao,2JKBhattacharjee1,2JKBhattacharjee2,rev_sengers}.
Like the static case, the dynamic exponents are also not all independent of each other, they follow 
certain scaling relations. E.g. starting from the generalized Stokes-Einstein-Sutherland relation
\cite{2HEStanley,2AOnuki1,2JLStrathmann,2JPHansen}
\begin{equation}\label{chap2_SES_rel}
D=\frac{R_D k_B T}{6\pi \eta \xi},
\end{equation}
$k_B$ being the Boltzmann constant and $R_D$ another universal constant \cite{2JLStrathmann}, one 
obtains \cite{2HEStanley}
\begin{equation}\label{chap2_dyna_scl_rel}
x_{_{D}}=1+x_{_{\eta}}.
\end{equation}
\par
~ Unlike the static case, the computational estimation
of the dynamic critical exponents started only recently, as mention above.
In this work, we have presented simulation results for the critical dynamics of a three dimensional single 
component Lennard-Jones (LJ) fluid that exhibits vapor-liquid transition. We 
focus on the bulk viscosity and the thermal conductivity. 
There, of course, exist simulation studies on dynamics in vapor-liquid transitions \cite{2KMeier,2AChen,2KDyer,Sear1}. 
In fact, in some previous studies \cite{2KMeier,Sear1} both these transport properties
were calculated in the vicinity of critical points. However, presumably due to computational difficulty with respect to the
calculation of collective transport properties, corresponding critical exponents  
were not quantified in those \cite{2KMeier,Sear1} works. On the other hand, even though 
the critical behavior of the thermal diffusion constant was studied in Ref. \cite{2AChen}, the associated conductivity  
was not separately looked at.
\par
~For this purpose, we have performed molecular dynamics (MD) simulations 
and analyzed the results via appropriate application of the finite-size scaling (FSS) theory \cite{2MEFisher4}. 
Prior to that, we have studied the phase behavior of the model by using the Gibbs ensemble Monte Carlo (GEMC) simulation 
method \cite{2AZPanagiotopoulos} as well as successive umbrella sampling technique \cite{Virnau1}
in $NPT$ ensemble \cite{Wilding1,Panagio1} ($N$ and $P$ are the total number of particles and pressure, respectively).
The critical temperature ($T_c$) and critical density ($\rho_c$) were estimated accurately via appropriate
FSS analyses \cite{Wilding2,Kim1,Claud1}.
\par
~ The rest of the paper has been organized as follows. In section II we have discussed the model and methodologies. 
The results are presented in section III. Finally, in section IV we have summarized our results.
\section{Model and Methods}
~ As stated, we have considered a single component LJ fluid. In our model, a pair of particles, $i$ and $j$, separated 
by a distance $r$ ($=|\vec{r}_i -\vec{r}_j|$), interact via the potential \cite{2MPAllen}
\begin{eqnarray}\label{chap2_LJ_pten0}
 U(r) &=& u(r)-u(r_c)-(r-r_c)\frac{du}{dr}\Big{|}_{r=r_c},~~ \mbox{for}~r\le r_c\nonumber \\
      &=& 0, ~~ \mbox{for}~r > r_c, 
\end{eqnarray}
where $r_c$ ($=2.5\sigma, \sigma$ being the particle diameter) is a cut-off distance, introduced to accelerate the computation.
In Eq. (\ref{chap2_LJ_pten0}), $u(r)$ is the standard LJ potential \cite{2MPAllen,2DFrankel}
\begin{equation}\label{chap2_LJ_pten}
 u(r)=4\varepsilon \Big[\Big(\frac{\sigma}{r}\Big)^{12}-\Big(\frac{\sigma}{r}\Big)^6\Big],
\end{equation}
with $\varepsilon$ being the interaction strength. For the sake of convenience we set $\sigma$ and $\varepsilon$ to unity.
The last term in the first part of Eq. (\ref{chap2_LJ_pten0}) was introduced to correct for the discontinuity in the force 
at $r=r_c$ that occurs after the cutting and shifting of the potential.
\par
~ The  GEMC simulations \cite{2DFrankel, 2AZPanagiotopoulos}, for the study of the phase behavior of the model, were performed 
in two separate boxes, as discussed below. The total number of particles in and the total volume ($V$) 
of the two boxes were kept fixed, though the numbers of particles ($N_1$ and $N_2$) in as well as the volumes ($V_1$ and $V_2$)
of the individual boxes were varied during the simulations. 
We considered three types of perturbations or 
trial moves, viz., particle displacement in each of the boxes, volume change of the individual boxes and  
particle transfer between the boxes. Thus, this is a combination of simulations in constant $NVT$, $NPT$ and
$\mu_c VT$ ensembles, $\mu_c$ being the chemical potential.
At a late time, one observes coexistence of the vapor phase (in one of the boxes) 
with the liquid phase (in the other box), if a simulation is performed at a temperature $T< T_c$. 
Thus, by running the simulations at different temperatures and obtaining the equilibrium densities 
($\rho_{\alpha}=N_{\alpha}/V_{\alpha}$, $\alpha$ standing for liquid or vapor) of the individual phases, the whole phase diagram can be drawn, which, 
of course, will provide information about the critical temperature and critical density.
\par
The phase diagram was also obtained via successive umbrella sampling \cite{Virnau1} MC simulations
in $NPT$ ensemble \cite{Wilding1,Panagio1}. Like the grandcanonical case, the overall density fluctuates in this
ensemble as well. While in the former the fluctuation is a result of particle addition and deletion moves, in the case of
$NPT$ simulations the volume moves give rise to the fluctuation. The $NPT$ ensemble has advantage over the former
when overall density is rather high. In the implementation of successive umbrella sampling technique, for 
overall density $\in [0,1]$,
the corresponding volume range is divided into small windows. In each of these windows simulations were performed
over long periods of time. For $T<T_c$, these simulations provide
double-peak distribution for specific volume $v_{\rm sp}$ ($=V/N$). The peak at the smaller value of $v_{\rm sp}$,
at a particular temperature, corresponds to a point on the
liquid branch of the coexistence curve. The coexisting vapor density is given by the position
of the peak at the higher value of $v_{\rm sp}$. 
While the coexistence curve data will be presented from the GEMC simulations, for
the estimation of critical parameters, particularly $\rho_c$, we will rely on the simulations in $NPT$ ensemble. 
Here note that our results on the phase
behavior are consistent with the data from the simulations in 
grandcanonical ensemble which are made available online \cite{NIST1}.
\par
~ To study the transport properties we have performed MD simulations \cite{2MPAllen,2DFrankel,2DCRapaport}. There we first 
thermalize the systems, using the stochastic Andersen thermostat \cite{2DFrankel}, to generate the initial configurations.
Finally, for the production runs we performed MD simulations in the 
microcanonical (constant $NVE$, $E$ being the total energy) ensemble that preserves hydrodynamics, 
essential for the calculations of transports in fluids \cite{2DFrankel}.
\par
~ The transport quantities have been calculated by using the Green-Kubo (GK) formulae \cite{2JPHansen,2MPAllen}. 
The GK relations for the viscosities and the thermal conductivity are connected to the expressions \cite{2JPHansen,2MPAllen}
\begin{equation}
\mathcal{Y}=\frac{1}{k_{B}T V} \int_{0}^{t} dt' <\sigma'_{\mu s}(t') \sigma'_{\mu s}(0) >; ~\mu,s \in[x,y,z],
\label{chap2_GK_vist}
\end{equation}
and
\begin{equation}
\lambda=\frac{1}{k_{B}T^2 V} \int_{0}^{t} dt' <j^{s}_{T}(t) j^{s}_{T}(0) >; ~s \in[x,y,z].
\label{chap2_GK_convty}
\end{equation}
In Eq. (\ref{chap2_GK_vist}), $\sigma'_{\mu s}$ is related to the pressure tensor $\sigma_{\mu s}$, defined as
\begin{equation}\label{chap2_pt_bvist}
\sigma_{\mu s}=\sum_{i=1}^{N} \Big[m_i v_{i\mu} v_{i s} +\sum_{j=i+1}^{N} (\mu_i - \mu_j) F_{s j}\Big], 
\end{equation}
where $F_{sj}$ is the $s^{\mbox{th}}$ component of the force on the $j^{\mbox{th}}$ particle, $m_i$ is the mass of the $i^{\mbox{th}}$
particle (chosen to be equal to $m$ for all), $v_{i\mu(s)}$ is the $\mu (s)^{\mbox{th}}$ component of velocity for particle $i$
and $\mu_{i(j)}$ is the Cartesian  coordinate for particle $i$($j$) along the $\mu$-axis.
For the diagonal elements $\sigma'_{\mu \mu}=\sigma_{\mu \mu}-<\sigma_{\mu \mu}>$ and  $\mathcal{Y}=\zeta+4/3\eta$, whereas  
for the off-diagonal elements ($\sigma'_{\mu s}=\sigma_{\mu s}$) $\mathcal{Y}=\eta$. In Eq. (\ref{chap2_GK_convty}), $j_{T}^{s}$
is the thermal flux along any particular axis, defined as 
\begin{equation} \label{chap2_flux_tc}
j_{T}^{s}=\frac{1}{2}\sum_{i=1}^{N} v_{is}\Big[ m |v_i|^2 +\sum_{j\ne i}^{N} U(r)\Big]
-\frac{1}{2} \sum_{i=1}^{N} \sum_{j\ne i}^{N} \vec{v}_{i}\cdot \vec{r}\frac{\partial U(r)}{\partial s},
\end{equation}
where $v_{is}$ is the velocity component of the $i^{\mbox{th}}$ particle along $s$-axis. In $U(r)$ it is understood that the energy 
comes from the interaction between particles $i$ and $j$, vector distance between them being
represented by $\vec{r}$. This justifies the summation over $j$ in the last equation.
\par
~ All our simulations were performed in cubic systems of linear dimension $L$ and in the presence of periodic boundary conditions 
in all possible directions. In our MD simulations, time was measured in an LJ unit $t_0$ ($=\sqrt{m \sigma^2/\epsilon}$)
and the integration time step was set to $dt=0.005 t_0$. All the results related to 
transport properties are presented after averaging over 
64 initial realizations. From here on, for the sake of convenience, we set $m$, $k_B$ and $t_0$ to unity.
Note that the time in MC simulations is expressed in units of number of Monte Carlo steps (MCS).
In the case of GEMC method, each step consists of 
$75\%$ displacement moves, $10\%$ volume moves, and $15\%$ particle transfer moves, of a total of $N$ trials.
There was no particular order for the execution of these moves.
Results for the coexistence curve are presented after averaging over 15 initial configurations.
\section{Results}
\subsection*{A. Phase Behavior}
\begin{figure}[h!]
\centering
\includegraphics*[width=0.4\textwidth]{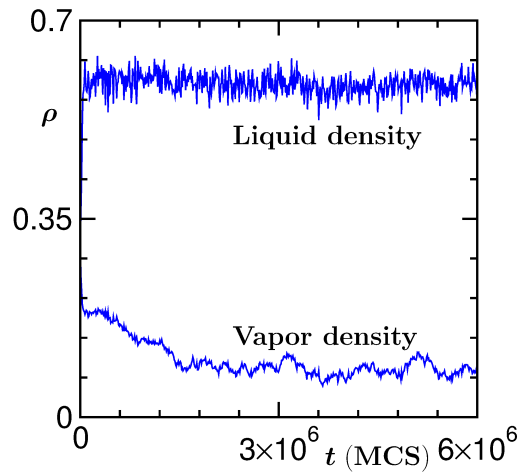}
\caption{\label{fig1} Density profiles inside the two boxes, during a Gibbs ensemble Monte Carlo 
run with $V=2\times12^3$, are plotted vs time. The results correspond to $T=0.86$.}
\end{figure}

\begin{figure}[h!]
\centering
\includegraphics*[width=0.4\textwidth]{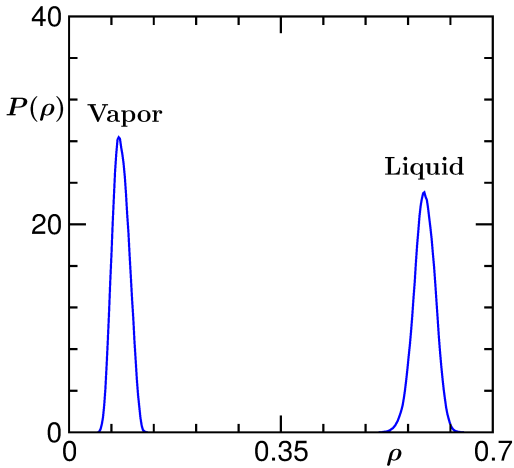}
\caption{\label{fig2} Plot of the density distribution function, $P(\rho)$, vs $\rho$, for the density profiles in Fig. \ref{fig1}.}
\end{figure}

~ In Fig. \ref{fig1} we show the density profiles inside the two boxes, vs time, obtained from a typical run in the GEMC 
simulations \cite{2AZPanagiotopoulos} at $T=0.86$. For each of the studied temperatures, we started with density $\rho=0.3$, 
in each of the boxes. Gradually, the density in one of the boxes increases with time, 
while it decreases in the other box, if $T<T_c$. Finally, the densities inside both the boxes saturate and fluctuate
around the mean values, as shown in this figure. The distribution of the densities, obtained from the profiles in Fig. \ref{fig1}, 
has been presented in Fig. \ref{fig2}. The appearance of the two peaks is expected 
(given that the profiles are well separated)  
and implies the coexistence of vapor and liquid phases. There the locations of the peaks correspond to the 
equilibrium density values of the vapor and liquid phases, for the studied temperature.
\par
\begin{figure}[h!]
\centering
\includegraphics*[width=0.4\textwidth]{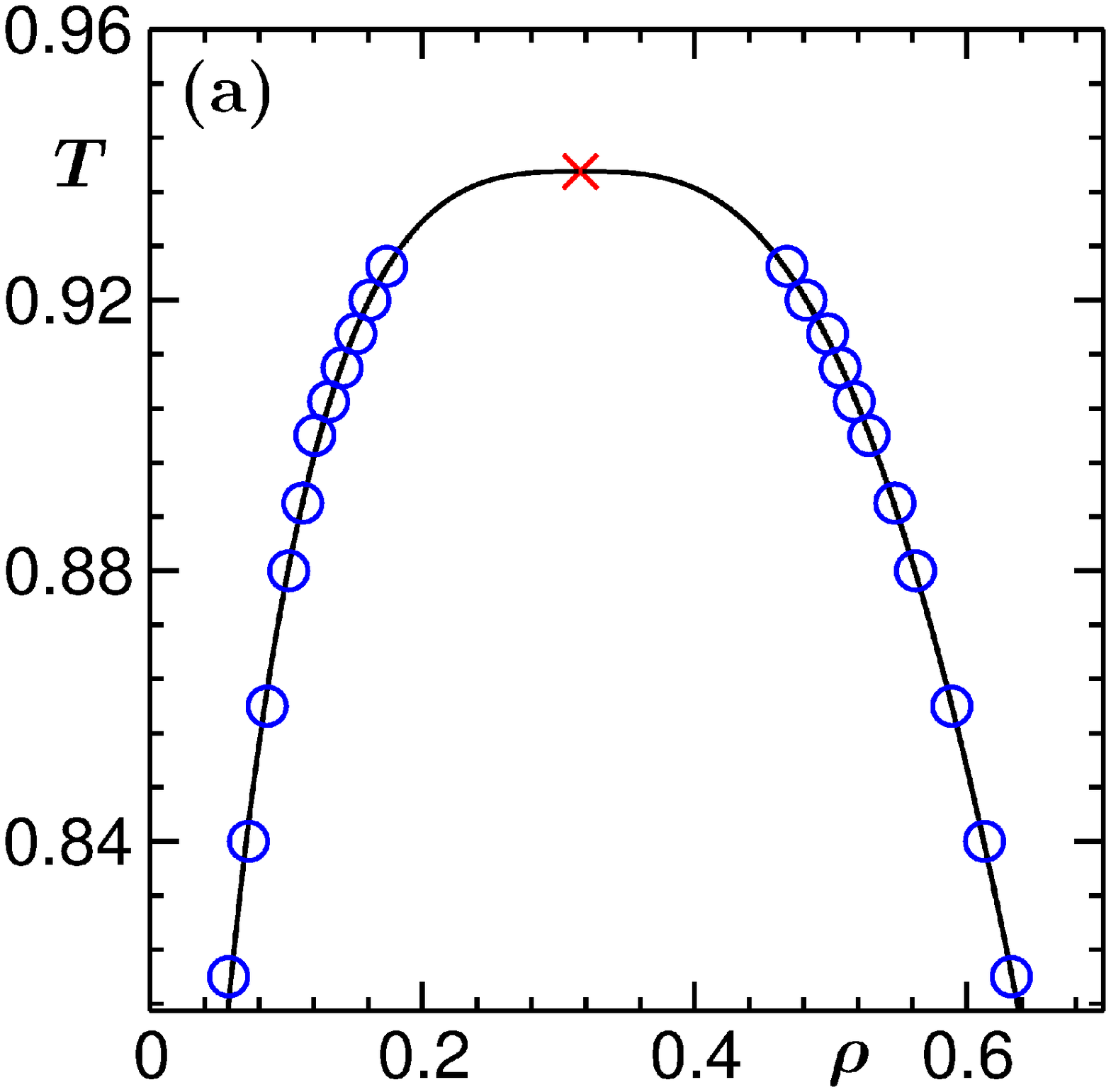}
\includegraphics*[width=0.4\textwidth]{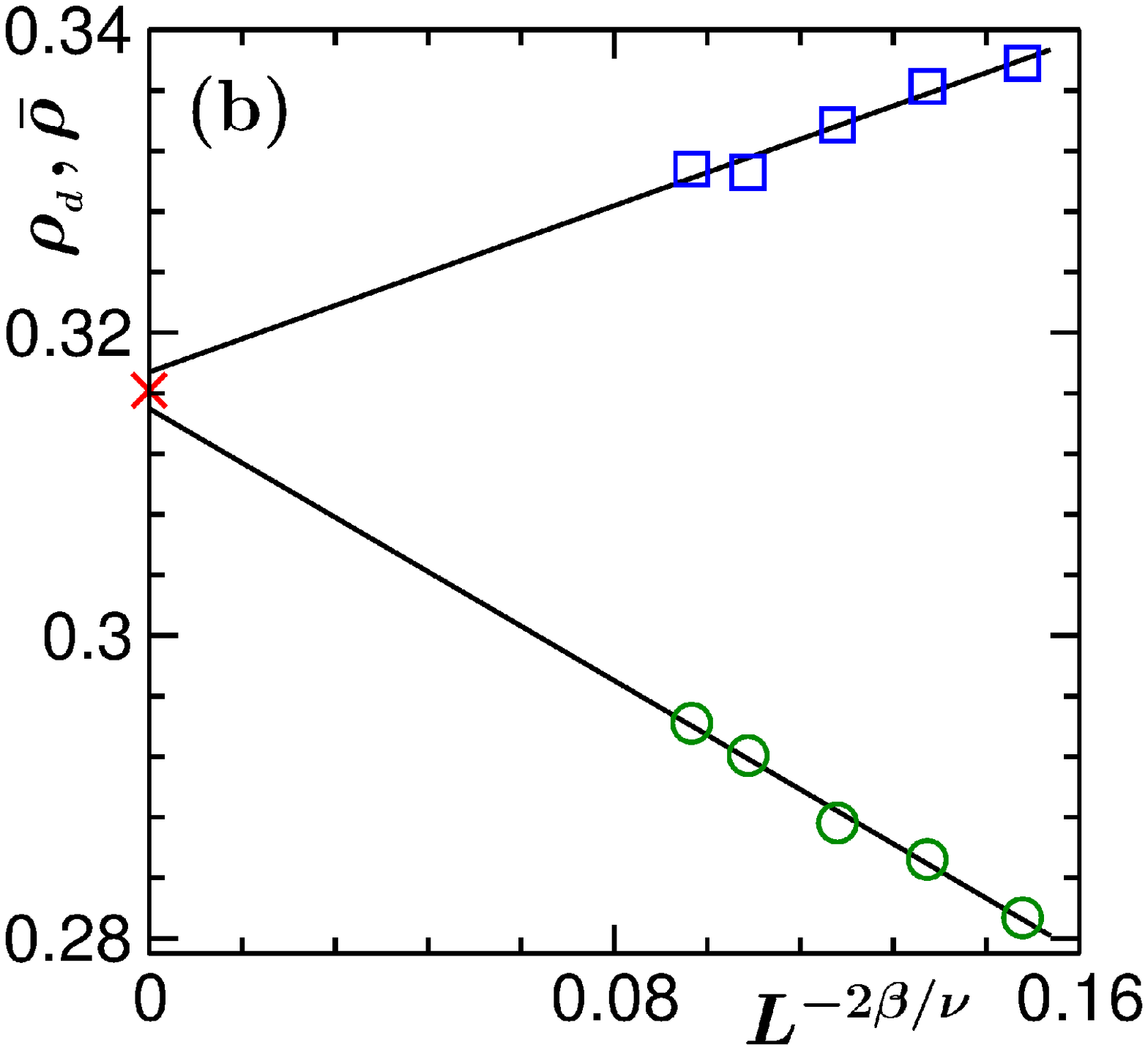}
\caption{\label{fig3} (a) Phase diagram of the 3D LJ fluid in the $T-\rho$ plane, obtained via the Gibbs ensemble 
Monte Carlo simulations. The cross mark in the figure is the location of the critical point. The continuous line represents 
the Ising critical behavior of the order parameter. The results correspond to $V=2\times12^3$.
(b) Demonstration of the estimation of $\rho_c$ via finite-size scaling analysis. Here we have plotted
$\rho_d$ (upper curve) and $\bar{\rho}$ (lower curve), obtained from $NPT$ simulations at $T_c$, vs $L^{-2\beta/\nu}$}.
\end{figure}

~ In Fig. \ref{fig3} (a) we have presented the phase diagram for the model, in the temperature vs density plane. 
We obtained this by plotting the equilibrium coexistence densities of the two phases at different temperatures. 
Accuracy of these 
results are checked by comparing with the ones obtained from  
umbrella sampling simulations in the $NPT$ ensemble. 
From this 
figure, it is clear that the value of the order-parameter $\psi$ ($=\rho_\ell - \rho_v$, $\rho_\ell$ and $\rho_v$ being 
respectively the liquid and vapor densities) is approaching zero with the increase of temperature. In Fig. \ref{fig3} (a), we do not 
present data from temperatures very close to critical point, since they suffer from the finite-size effects. 
The finite-size effects were appropriately identified by comparing the results from different system sizes. 
\par
~ The values of $T_c$ and $\rho_c$ can be calculated by using the equations \cite{2DFrankel}
\begin{equation}\label{op_vs_epsi}
\psi=\rho_\ell -\rho_v =A(T-T_c)^\beta,
\end{equation}
and
\begin{equation}\label{aop_vs_epsi}
 \rho_d=\frac{\rho_\ell+\rho_v}{2}=\rho_{c} + B(T-T_c),
\end{equation}
where $A$ and $B$ are constants. For fitting the simulation 
data to Eq. (\ref{op_vs_epsi}), to obtain $T_c$, we choose $\beta=0.325$, which, as already mentioned, is its value 
for the $d=3$ Ising universality class. Since LJ potential is a short-range one, this value is expected. For the same reason, 
we will adopt the Ising value for $\nu$, while analyzing the transport properties. 
This exercise provides $T_c=0.939\pm0.004$. This is in good agreement with
a previous estimate via grandcanonical simulations, for the same model \cite{Errin1}.
\par
Estimation of $\rho_c$, on the other hand, will suffer from error, 
if made via fitting to Eq. (\ref{aop_vs_epsi}). This
is because, Eq. (\ref{aop_vs_epsi}) should contain additional terms 
in powers of ($T_c-T$), due to field mixing \cite{Wilding2,Kim1,Claud1}.
Accurate finite-size scaling analyses \cite{Wilding2,Kim1} 
have been performed to extract $\rho_c$, that take care of these singularities. In 
some of these previous studies \cite{Wilding2,Errin1} only the
term proportional to $\epsilon^{1-\alpha}$ have been considered. More recently, it has been
stressed that the leading singularity \cite{Kim1,Claud1} is $\epsilon^{2\beta}$ and should be considered
for more accurate estimation of $\rho_c$. Here we perform finite-size scaling
analysis using this dominant contribution. For this exercise we have used data from NPT simulations
at $T_c$. Recall that, like $L$ in the grandcanonical ensemble, here $N$ is kept fixed and we treat it
as $L^3$.
\par
In Fig. \ref{fig3} (b) we show $\rho_d$ (upper curve) as a function of $L^{-2\beta/\nu(=1.032)}$. 
This scaling form comes from the fact that $\xi\sim L$ at $T_c$. Linear extrapolation
of the data set to $L=\infty$ provides $\rho_c\simeq 0.317$. In this figure
we have also included the mean value of $\rho$ ($\bar{\rho}$) (see lower plot), estimated from the inverse of the average specific 
volume. This also exhibits a linear behavior, extrapolation of which leads to $\rho_c\simeq 0.315$.
From these exercises we take $\rho_c=0.316$. 
In Fig. \ref{fig3} (a), the cross mark is the location of the critical point.
The simulation data in this figure show nice consistency with the continuous line, which has the Ising behavior.
Our estimation of $\rho_c$ is reasonably consistent with the previous \cite{Errin1} grandcanonical
estimate ($0.320$). Little more than $1\%$ difference that exists may well be due to the fact that 
in this earlier work data were not analyzed by considering the leading singularity. 
Nevertheless, in view of this difference,
we have calculated transport properties over a wide range of density, viz. $[0.31,0.32]$. While we will
present results at our estimated value of $\rho_c$, outcomes from other densities will be mentioned in 
appropriate place.
\par
Note that the values of $T_c$ and $\rho_c$ were estimated previously \cite{2DFrankel,Wilding2} for the vapor-liquid 
transitions in similar LJ models. However, those studies either used different values of $r_c$ or did
not consider the term related to force correction. The difference in the numbers between our study and these
previous ones are related to these facts. In fact, the cut-off dependence of the critical temperature is
nicely demonstrated by Trokhymchuk and Alejandre \cite{Trokh1}. However, we cannot use the information
from this work because of the force correction that we use.
\par
\begin{figure}[h!]
\centering
\includegraphics*[width=0.4\textwidth]{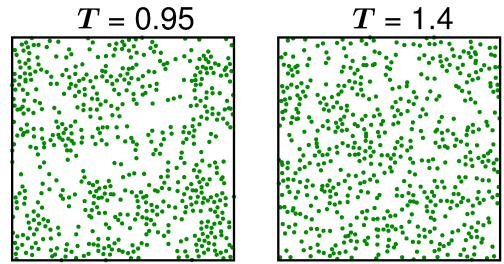}
\caption{\label{fig4} Two-dimensional slices of typical equilibrium configurations at $T=0.95$ and $1.4$. 
The dots mark the locations of the particles.}
\end{figure}

~ Before proceeding to show the results for dynamics, in Fig. \ref{fig4} we show the two-dimensional cross-sections of two typical 
equilibrium configurations at $T=0.95$ and $1.4$. Structural difference between the two snapshots is clearly visible. The one at $T=0.95$ 
shows density fluctuations at much larger length scale, implying critical enhancement in $\xi$. 
The values of $\xi$, as well as $\chi$, can be calculated 
from the density-density structure factors by 
fitting the small wave-vector data to the Ornstein-Zernike form \cite{2HEStanley}.
\subsection*{B. Dynamics}
\begin{figure}[h!]
\centering
\includegraphics*[width=0.4\textwidth]{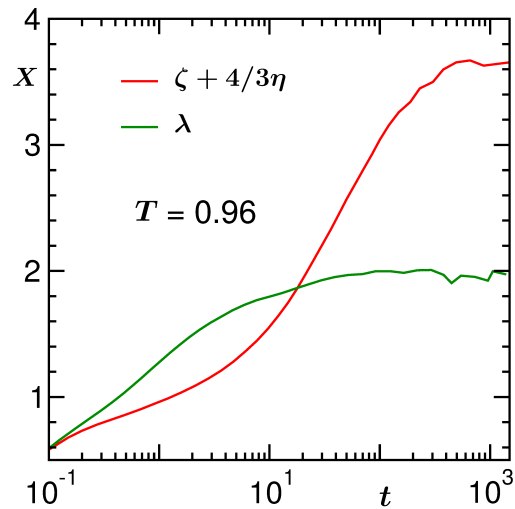}
\caption{\label{fig5} Plots of $X$ ($\zeta+\frac{4}{3}\eta, \lambda$) vs $t$, in a semi-log scale, at $T=0.96$, with $L=30$.}
\end{figure}

~ All the results for dynamics are presented from temperatures above the critical value, by fixing $\rho$ to $\rho_c$.
In Fig. \ref{fig5}, we show the plots of $\zeta+\frac{4}{3}\eta$ and $\lambda$, vs time, as obtained from the GK formulas, 
at $T=0.96$, on a semi-log scale. We extract the final values for these quantities from the flat regions. 
From this figure it is clear that a transport quantity having higher critical exponent settles down to a flat plateau 
at a later time. This states about 
the difficulty of calculating a transport coefficient with strong critical divergence, 
like the bulk viscosity ($\zeta$), particularly close to $T_c$. The difficulty gets pronounced with the increase of system size, 
consideration of which is essential to avoid the finite-size effects in the critical vicinity. However, in our simulations we have 
used relatively small system sizes and relied on the FSS theory \cite{2MEFisher4} for the estimation of
the critical exponents. 
\par
\begin{figure}[h!]
\centering
\includegraphics*[width=0.4\textwidth]{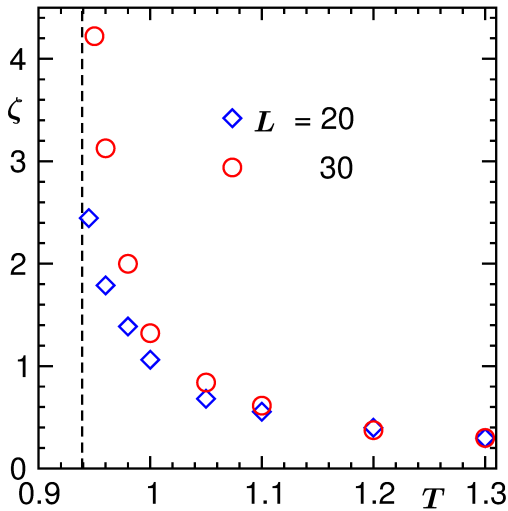}
\caption{\label{fig6} Plots of $\zeta$ vs $T$ for two different system sizes. Close to the critical point the error 
bars are of the order of the symbol sizes. The dashed line in the figure is the location of the critical temperature.}
\end{figure}
\begin{figure}[h!]
\centering
\includegraphics*[width=0.4\textwidth]{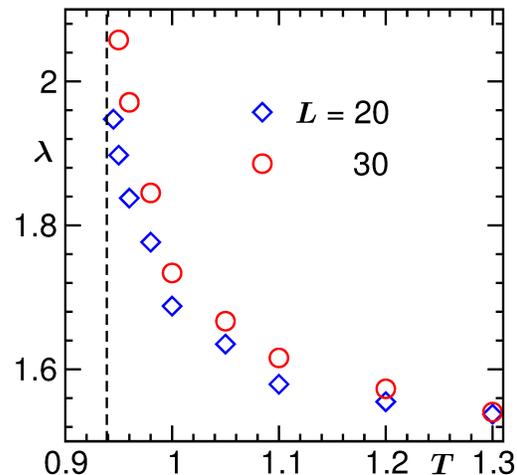}
\caption{\label{fig7} Plots of $\lambda$ vs $T$. Data from 
two different system sizes are shown. Close to the critical point the error bars are of the order of the symbol sizes. 
The dashed line marks the location of the critical temperature.}
\end{figure}

~ The temperature dependence of the bulk viscosity and the thermal conductivity, obtained from the plateaus of GK integrations,
have been presented in Fig. \ref{fig6} and Fig. \ref{fig7}, respectively. The enhancement in these quantities can be 
observed for both the presented system sizes, mentioned in the figure, close to $T_c$, represented by the dashed lines. 
Weaker enhancement for the smaller system, for both $\zeta$ and $\lambda$, signify finite-size effects. 
\par
\begin{figure}[h!]
\centering
\includegraphics*[width=0.4\textwidth]{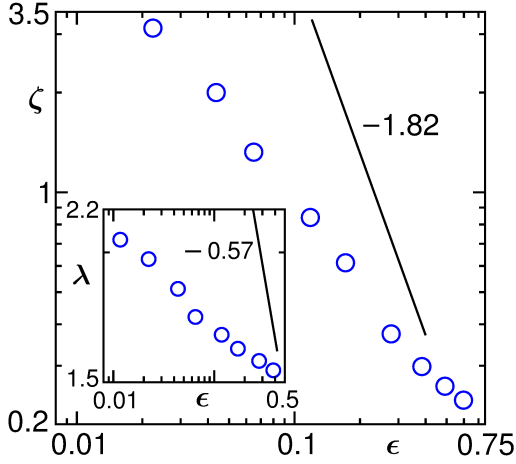}
\caption{\label{fig8} Plot of $\zeta$ vs $\epsilon$, on a log-log scale, for $L=30$. The solid line
corresponds to the theoretical expectation. Inset shows the same exercise for $\lambda$.}
\end{figure}

~ In Fig. \ref{fig8} we show the plot of $\zeta$ vs $\epsilon$, using data from the larger system size that has been 
used in Fig. \ref{fig6}, on a log-log scale. We observe that the simulation data are in disagreement with the 
theoretically predicted solid line (having exponent $x_{_{\zeta}}\nu=1.82$). 
The reasons for the disagreement could be the finite-size effects as well as the presence of a 
background contribution \cite{2HCBurstyn3}, 
the latter arising from small wavelength fluctuations. We observe similar disagreement for $\lambda$, 
presented in the inset of Fig. \ref{fig8}, for the same system size. These two serious issues, viz., 
finite-size effects and background contributions, have to be appropriately taken care of during the estimation 
of the critical exponents, along the line discussed below. 
\par
~ A quantity, say $X$, that exhibits singularity at the critical point, can be decomposed into two 
parts \cite{2JLStrathmann,2SKDas1,2SKDas2,2HCBurstyn3} as
\begin{equation} \label{chap2_decom_X}
X=\Delta X(T) + X_{_b} ,
\end{equation}
where $\Delta X(T)$ comes from the critical fluctuations and
is strongly temperature dependent. On the other hand, $X_b$, the background,
is only weakly temperature dependent and is often treated as a constant \cite{2SKDas1,2SKDas2}.
This latter contribution should also be independent of the system size. The presence of such a term, particularly 
in computer simulations, where one works with finite systems, can lead to a misleading conclusion. 
To extract the correct critical divergence one needs to subtract it appropriately from the total value, such that  
\begin{equation} \label{chap2_cri_beh_dX}
\Delta X(T) =X-X_b \sim \xi^{x}, 
\end{equation}
where $x$ is the critical exponent for $X$. We have estimated $X_b$ by treating it
as an adjustable parameter in the FSS analysis that we describe below. One might as well have aimed to
obtain the background contributions from Fig. \ref{fig8} by looking at the behavior of the data sets far away from $T_c$.
Even though these plots certainly provide hint on the presence of nonzero $X_b$, even a weak 
temperature dependence of the latter may cause significant error while analyzing data 
close to $T_c$, if estimated from high $T$ convergence.
\par
~ As stated above, at the critical 
point the correlation length is restricted by the system size, i.e., $\xi \sim L$ at $T=T_c$, so that \cite{2DPLandau}
\begin{equation} \label{chap2_X_vs_L}
 \Delta X(T_c) \sim L^{x}.
\end{equation}
Far from $T_c$, the finite-size effects will be absent, i.e., the data will be independent of $L$. 
To describe the thermodynamic limit 
($L\gg\xi$) and finite-size limit data by a single equation, one should introduce a bridging or FSS function $Y(y)$, to write  
\begin{equation} \label{chap2_int_scl_func}
 \Delta X(T) \sim Y(y) L^{x}.
\end{equation}
In Eq. (\ref{chap2_int_scl_func}), $Y(y)$ is independent of the system size and depends upon the scaling 
variable $y$ ($=(L/\xi)^{1/\nu} \sim \epsilon L^{1/\nu}$), the latter being a dimensionless quantity. 
In the limit $y\rightarrow 0$, i.e., $T\rightarrow T_c$, $Y$ must be a constant so that Eq. (\ref{chap2_X_vs_L}) is recovered.
On the other hand, in the limit $y\rightarrow \infty$ ($\xi<<L, ~\epsilon \gg 0$), 
$Y$ should exhibit a power-law decay  
\begin{equation} \label{chap2_Y_vs_y}
 Y(y) \sim  y^{-x\nu},
\end{equation}
so that the data are described 
by Eq. (\ref{chap2_cri_beh_dX}). A plot of $Y$ vs $y$, obtained by taking data from different system sizes, 
will exhibit data collapse, for appropriate choices of $X_b$, $x$ and $\nu$. Also, for the best data collapse, 
the large $y$ behavior of $Y$ will be consistent with Eq. (\ref{chap2_Y_vs_y}).
\begin{figure}[h!]
\centering
\includegraphics*[width=0.4\textwidth]{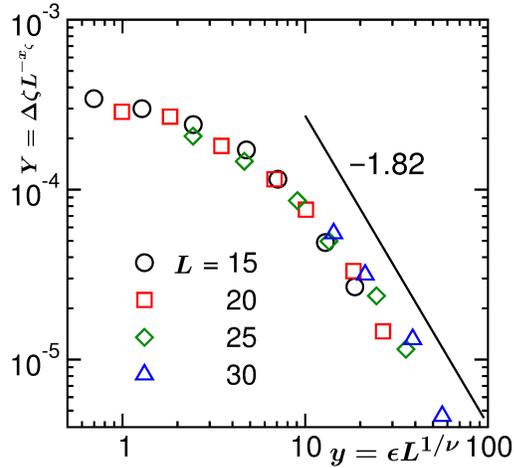}
\caption{\label{fig9} Finite-size scaling plot for the bulk viscosity.
The scaling function $Y$ ($=\Delta \zeta L^{-x_{_{\zeta}}}$) is plotted vs the scaling parameter
$y$ ($=\epsilon L^{1/\nu}$), on a log-log scale, using data from different system sizes.
The solid line in the figure represents a power-law with the exponent being mentioned next to it. }
\end{figure}
\begin{figure}[h!]
\centering
\includegraphics*[width=0.4\textwidth]{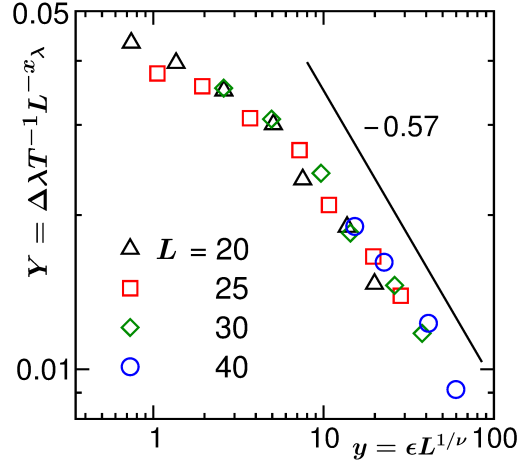}
\caption{\label{fig10} Finite-size scaling exercise for the thermal conductivity. Here we show
$Y$ ($=\Delta \lambda T^{-1} L^{-x_{_{\lambda}}}$) vs $y$ ($=\epsilon L^{1/\nu}$) on a
log-log scale.
The solid line is a power-law, exponent of which is mentioned next to the line.}
\end{figure}
\begin{figure}[h!]
\centering
\includegraphics*[width=0.4\textwidth]{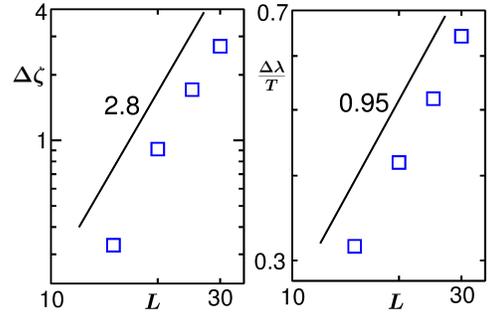}
\caption{\label{fig11} Log-log plots of critical parts of bulk viscosity (left) and thermal conductivity (right),
vs $L$, at $T_c^{\rm eff}(f,L)$ with $f=1$. The solid lines are power laws. Corresponding exponent values
are mentioned.}
\end{figure}
\par
~ In Fig. \ref{fig9}, we have presented the FSS analysis result for $\zeta$, by plotting $Y(y)$ vs $y$, using data from 
different system sizes, mentioned on the figure. To show consistency with the theoretical predictions, 
in this analysis we have used $\zeta_b$ (background contribution for $\zeta$)
as adjustable parameter and fixed $\nu$ and $x_{\zeta}$ to their theoretical values.
The presented result corresponds to best collapse which is obtained for $\zeta_b=0.40$. Given
the difficulty one encounters in calculating bulk viscosity, even a reasonably better 
collapse would require significant additional effort.
In the limit $y\rightarrow 0$, the master curve approaches a constant value, 
as expected from the construction of $Y$. On the other hand, for $y\rightarrow \infty$, the master curve is 
showing a power-law decay with the exponent $x_{\zeta}\nu=1.82$. 
Similar exercise we have performed for $\lambda$, the results for which are presented in Fig. \ref{fig10}. 
Here note that, since $\Delta\lambda \sim T\epsilon^{-0.57}$, the ordinate contains 
the factor $T^{-1}$.
In this case we have obtained best collapse for
$\lambda_{_b}=1.34$. 
\par
To justify the correctness of the background values obtained above, we perform 
further analysis \cite{2SRoy1,DKF1,DKF2}. This, in addition to achieving the stated objective, will provide direct
information on the critical exponents as well. For this purpose, we define finite-size effective 
critical points as
\begin{equation}\label{tcl}
T_c^{\rm eff}(f,L)=T_c+f(T_c^L-T_c).
\end{equation}
Even though we do not have estimates of the finite-size critical points $T_c^L$, $T_c^{\rm eff}$
can be estimated from the fact \cite{2MEFisher4} that $(T_c^L-T_c)\sim L^{-1/\nu}$. Data at
$T_c^{\rm eff}(f,L)$, for various values of $f$, will have same scaling form
as that at $T_c^L$. Thus, we expect $\Delta X$ to behave as
$\Delta X \sim L^x$, when extracted at $T_c^{\rm eff}(f,L)$ for a fixed value of $f$.
In Fig. \ref{fig11} we have performed this exercise for both $\zeta$ and $\lambda$ for 
$f=5$. In this process we have subtracted the values of background that we obtained
above. The value of $f$ was chosen in such a way that the effective finite-size
critical points do not fall in the finite-size coexistence region and 
corresponding values of $\epsilon$ do not exceed $0.1$. Results at various values
of $T_c^{\rm eff}(f,L)$ were obtained by suitable interpolation using the existing temperature
dependent data for different values of $L$. These results are presented on log-log scales. 
The data are consistent with the theoretical expectations,
within about $5\%$ deviation. We could as well have estimated
the backgrounds from this exercise and used the numbers in the FSS analyses of Figs. \ref{fig9}
and \ref{fig10}.
\par
All the results on dynamics have been presented for $\rho=\rho_c$. As stated above, we have
accumulated data over a wide range of density. Similar FSS analyses have been 
performed for $\rho =0.31$ and $0.32$. For these values of $\rho$, we observe that the exponent values are in 
reasonable agreement with the ones for $\rho=\rho_c$.
Such small difference is consistent with the data presented in Ref. \cite{Sear1}.
In this latter work, over a density range of about $5\%$ on either side of $\rho_c$, the
thermal conductivity data showed quite flat behavior.
\section{Summary}
~ We have studied the phase behavior and the dynamic critical phenomena for vapor-liquid transition 
in a single component Lennard-Jones fluid in space dimension $d=3$. The phase behavior was obtained 
via Monte Carlo simulations \cite{2AZPanagiotopoulos}. 
To study the dynamic critical phenomena, we performed molecular dynamics simulations \cite{2MPAllen,2DFrankel,2DCRapaport} 
in microcanonical ensemble. The Green-Kubo relations \cite{2JPHansen} were used to calculate the transport  
quantities, viz., the bulk viscosity and the thermal conductivity. We observe strong finite-size effects,
similar to the case of liquid-liquid transitions \cite{2SKDas1,2SRoy1}. Our finite-size scaling analyses, 
however, show that the simulation data are consistent with the theoretically predicted critical divergences.
In fact, to the best of our knowledge, this is the first time the critical exponents for
bulk viscosity and thermal conductivity have been quantified for a vapor-liquid transition.

Our results, along with the ones for 
the binary fluid \cite{2SKDas1,2SRoy1}, are compatible with the expectation that the dynamic
critical phenomena of the vapor-liquid and liquid-liquid transitions belong to the 
same universality class, defined by model H \cite{2PCHohenberg}. Here note that the theoretical numbers for $x_{_{\zeta}}$
for vapor-liquid and liquid-liquid transitions are slightly different \cite{2JKBhattacharjee1,2JKBhattacharjee2}. 
This difference is within the error bars of computation via molecular dynamics.
\par
~ Despite the similar critical exponents in vapor-liquid and liquid-liquid transitions, we have observed some differences 
between the two cases. Our observation of the critical range in this work is less wide compared 
to that of the liquid-liquid transition \cite{2SKDas1,2SRoy1}. 
We also have observed that the background contribution for the bulk viscosity is nonzero (though small), 
whereas in the liquid-liquid transition it was not needed in the analysis \cite{2SRoy1}. Similarly, 
for thermal conductivity the background term plays very important role. These differences may have some connection with 
the symmetry of the model in the liquid-liquid case, but further investigations will be needed to confirm it. 
\par
~Acknowledgment: SKD and JM acknowledge financial supports from the Department of Science and Technology, Government of India,
and Marie Curie Actions plan of the European Union (FP7-PEOPLE-2013-IRSES Grant No. 612707, DIONICOS).
JM is grateful to the University Grants Commission, India, for research fellowship. The $NPT$ simulation
code was written with the objective of obtaining vapor-liquid coexistence curve in binary mixtures, in collaboration
with J. Horbach (JH). We thank JH for important inputs with respect to this.
\\
* das@jncasr.ac.in\\

\end{document}